\pdfoutput=1 

\documentclass{iaus}
\usepackage{graphicx}
\usepackage{amssymb}
\usepackage{epstopdf}
\title[Close pairs: keys to comprehension of the evolution of star clusters] 
{Close pairs: keys to comprehension the evolution of star clusters}

\author[Dany Vanbeveren]   
{Dany Vanbeveren$^{1.2}$}

\affiliation{$^1$ Astrophysical Institute, Vrije Universiteit Brussel  \\ email: {\tt dvbevere@vub.ac.be} \\[\affilskip]
$^2$ GroepT Leuven Engineering College, Association KU Leuven  \\email: {\tt dany.vanbeveren@groept.be}}

\pubyear{2008}
\volume{xxx}  
\pagerange{119--126}
\jname{Title of your IAU Symposium}
\editors{A.C. Editor, B.D. Editor \& C.E. Editor, eds.}
\begin{document}

\maketitle

\begin{abstract}
In this review I first summarize why binaries are key objects in the study of stellar populations, key objects to understand the evolution of star clusters, key objects to understand galaxies and thus the universe. I then focus on 4 specific topics: 

\medskip
\noindent 1. the formation (via binaries) and evolution of very massive stars in dense clusters and the importance of stellar wind mass loss. I discuss preliminary computations of wind mass loss rates of very massive stars performed with the Munich hydrodynamical code, and the influence of these new rates on the possible formation of an intermediate mass black hole in the cluster MGG 11 in M82

\medskip
\noindent 2. the evolution of intermediate mass binaries in a starburst with special emphasis on the variation of the SN Ia rate (the delayed time distribution of SN Ia). A comparison with SN ia rates in elliptical galaxies may provide important clues on the SN Ia model as well as on the evolution of the SN Ia progenitors

\medskip
\noindent 3. the evolution of the double neutron stars mergers in a starburst (the delayed time distribution of these mergers) and what this tells us about the suggestion that these mergers may be important production sites of r-process elements

\medskip
\noindent 4. the possible effect of massive binaries on the self-enrichment of globular clusters.

\keywords{Stellar dynamics, binaries: close, stars: evolution.}

\end{abstract}

\firstsection 
\section{Introduction}
\medskip
The question whether or not binaries are important in population studies in general, cluster studies in particular is obviously rhetoric but sometimes it is useful to summarize why the question is rhetoric. In section 2 I give a personal selection of observational and theoretical facts that illustrate the importance of binaries. In sections 3-5 I will highlight 3 topics related to binaries in clusters/starbursts:
\medskip
\begin{itemize}
\item one of the most spectacular events related to cluster stellar dynamics is the real physical collision of 2 or more stars. The most probable scenario for this collision process goes as follows: a primordial binary or a binary that is formed dynamically interacts with a third object (a single star or another binary); this may result in the formation of unconventionally formed objects = UFOs (Vanbeveren, 2007) where binary components are exchanged or where a new binary originates with a component which is a merger of 2 or more stars. Some standard massive binaries like the Wolf-Rayet binary $\gamma$$^2$ Velorum may have been formed this way (see also the latter paper). When the interaction results in a merger of two or more stars, it becomes more massive and attracts other stars. This may initiate a runaway collision process, which in turn may result in the formation of an intermediate mass black hole. The latter will be critically discussed in section 3
\medskip
\item the delay time distribution of supernovae Type Ia (SN Ia) in elliptical galaxies (considered as remnants of super starbursts) and what this tells us about the binary formation mechanism of these supernovae (section 4)
\medskip
\item the delay time distribution of merging double neutron star binaries (NS+NS) or merging neutron star + black hole binaries (NS+BH) and the link with the formation of r-process elements (section 5). 
\end{itemize}
\medskip

The present population of low mass stars in globular clusters (GCs) shows clearly the effects of chemical enrichment of a population of intermediate mass and/or massive stars that formed at an earlier evolutionary phase of the GC (see various contributions in the present proceedings). Studies aiming at explaining this self-enrichment have mainly focussed on single stars. In section 6 I give a few suggestions how massive binaries could have affected the chemical self-enrichment of globular clusters (see also the contribution in the present proceedings of Selma De Mink). 

\begin{figure}[b]
\begin{center}
\includegraphics[width=4in]{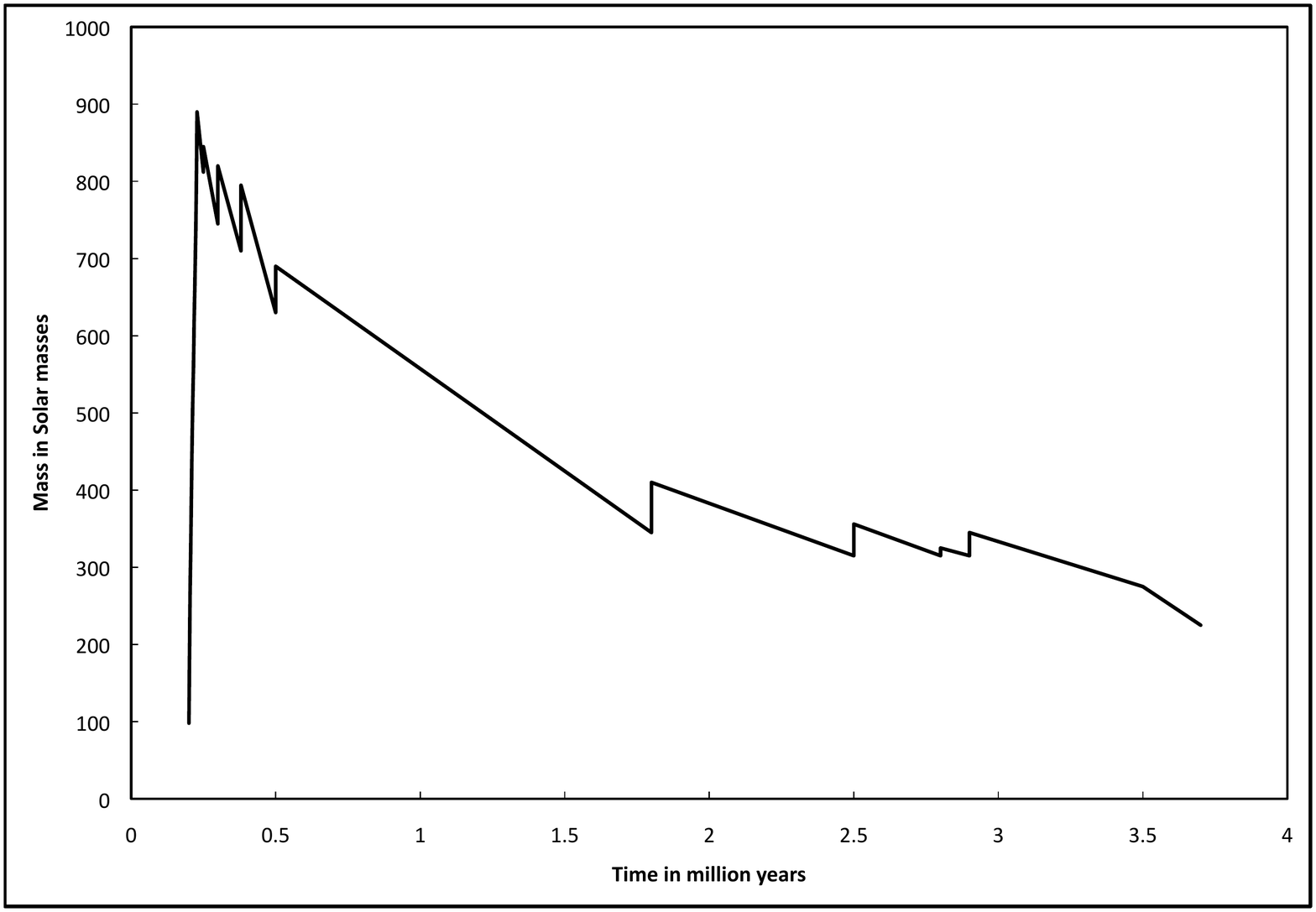} 
 \caption{The mass evolution of the runaway merger in the cluster MGG 11 predicted by an N-body code that includes in a self-consistent way life stellar evolution of massive and very massive stars, and where the stellar wind mass formalism of Pauldrach and Vanbeveren (2009) is used.}
\label{fig1}
\end{center}
\end{figure}

\section{The role of binaries in population synthesis}
\medskip
The influence of binaries on population number/spectral synthesis has been studied in numerous papers the last 3 decades (e.g., Van Bever and Vanbeveren, 2000, 2003;  De Donder and Vanbeveren, 2004; Belczynski et al., 2008 and references therein). It is clear that a discussion on the evolution of binaries is essential here, especially on uncertainties in all physical processes that govern binary evolution and how they affect population predictions.  Let me mention stellar wind mass loss during core hydrogen burning and core helium burning of massive stars, rotation, Roche lobe overflow and mass and angular momentum accretion in Case A and Case Br binaries, common envelope evolution in Case Bc and Case C binaries, the spiral-in process in binaries with an extreme mass ratio, the distribution of kick velocities of a compact SN remnant, and last not least, the evolution of binary mergers. I discussed some of these uncertainties in a review that has been published very recently (Vanbeveren, 2009). Below I list a number of facts resulting from all these studies (a personal and therefore probably a somewhat subjective selection).
\medskip
\begin{itemize}
\item Many (most) of the massive and intermediate mass stars in clusters are binary members (Mason et al., 2009; Kouwenhoven, 2006) but the binary frequency may vary from cluster to cluster, possibly depending on the cluster star density. About half of these binaries are interacting binaries (orbital period smaller than 10 years).  Note that the binary frequency among solar-type stars may be smaller (Zinnecker and Yorke, 2007). 
\medskip
\item During the evolution of a binary both components may merge and form a single star. When the primary of a massive binary explodes, due to the asymmetry of the SN explosion most of the binaries are disrupted. Due to stellar dynamics in dense clusters binaries may be disrupted. Although binaries may also be formed due to N-body processes in dense cluster, the net result of all this is that the presently observed binary frequency (at least in the massive and intermediate mass range) is smaller than the binary frequency at birth.
This also means that a fraction of the observed single stars may have had a binary evolutionary past. To illustrate, the single star $\zeta$ Pup in the Solar Neighborhood is a runaway which means that it most probably have had a binary history. In the past $\zeta$ Pup has frequently been considered as a prototype massive single star, but it is probably not a typical massive single star at all.
\medskip
\item About 10\% of the O-type stars are runaway stars (defined as stars with a peculiar space velocity $\ge$ 30 km/s) (Gies, 1987). We know of two processes able to produce runaway stars: close encounters in dense clusters of a binary and another object (either a single star or another  binary) (Lada et al., 1984) and the SN explosion in massive close binaries (Blaauw, 1961). Interestingly, in the close encounter scenario, many runaway stars are expected to be mergers of at least two stars (see for example the close encounter scenario for $\zeta$ Pup in Vanbeveren et al., 2009). At present there are too many unknown (cluster) parameters in order to determine the frequency of runaways formed via the close encounter scenario. De Donder et al. (1997) proceeded as follows: massive close binaries are a fact, SN explosions in these binaries are a fact and thus runaways formed through the Blaauw scenario are a fact. Using a binary population synthesis code it is possible to predict this frequency. We concluded that $\sim$5-8 \% of the O-type stars are expected to be Blaauw-type runaways (less than 1/3 of them have a compact companion, either a NS or a BH). In other words, we expect that $\sim$50-80 \% of the O-type runaways are formed via the SN explosion in binaries, and this means that 20-50\% of the O-type runaways may be formed via the close encounter process. 
\medskip
\item The influence of recent stellar wind mass loss rate formalisms (which account for the wind inhomogeneities = clumping) of massive core hydrogen burning stars (pre-LBV) on massive star evolution is very moderate and the uncertainties in these formalisms do not imply large uncertainties in overall massive star population synthesis. However, the mass loss rates during a possible Luminous Blue Variable phase, the Red Supergiant phase and the Wolf-Rayet phase (= the hydrogen deficient core helium burning phase of a massive star) are very important for stellar evolutionary prediction. The uncertainties in both the observed rates and the theoretically predicted ones is still a factor 2 or more and unfortunately this uncertainty critically affects massive star evolution and therefore also massive star population synthesis predictions.
\medskip
\item The rotational velocity distribution of O-type stars in the Solar Neighborhood reveals two important features: A. a significant fraction are relatively slow rotators with an average velocity of 100 km/s; when this is translated in an average initial velocity one arrives at the conclusion that a significant fraction of the O-type stars are born with an average velocity of $\sim$200 km/s. The evolutionary calculations of the Geneva group reveal then that the effect of rotation on the pre-SN explosion of this group of massive stars is rather moderate, comparable to the effect of moderate convective core overshooting (note that rotation implies rotational mixing of CNO elements up to the stellar surface, but this hardly affects the overall evolutionary behavior of a massive star);  B. the distribution has an extended tail towards rapid rotators, however a fraction of these rapid rotators are runaway stars indicating that they may have had a binary past. Mokiem et al. (2006) studied 21 OB dwarfs in the SMC and concluded that their average rotational velocity is $\sim$160-190 km/s. Since massive dwarfs are very young stars close to the zero age main sequence, the latter value should be representative for the average rotational velocity of massive stars at birth. Note then that this SMC value is very similar as the initial value for the galactic O stars given above whereas also in the SMC the most rapid rotators seem to be runaway stars, similar as in the Galaxy.
\medskip
\item	Rotating single star evolutionary models have difficulties to explain atmospheric CNO abundance anomalies in the observed massive star sample (Hunter et al., 2008). However, the combination of rotation and binaries gives a much better correspondence (Langer et al., 2008). I think that in this discussion the massive binary HD 163181 deserves some extra attention. It is an eclipsing binary with a period of 12 days, with a nitrogen enriched BN0.5Ia primary. Hutchings (1975) derived masses for the components, i.e. the primary mass = 13 M$_\odot$, the secondary mass = 22 M$_\odot$. The primary is 1.5-2 mag brighter than the secondary and this indicates that it is a core helium burning star that lost most of its hydrogen rich layers by Roche lobe overflow. This binary is therefore an illustration for the process where the atmospheric N-enhancement is due to a binary type mass loss rather than due to rotational mixing.
\medskip
\item	The observed overluminosity of the optical components of some of the standard massive X-ray binaries can be explained as due to rotational mixing in spun-up mass gainers of massive binaries (Vanbeveren and De Loore, 1994).
\medskip
\item	The effect of tides in short period binaries on the rotation of massive binary components in small metallicity regions (where stellar wind mass loss is small) explains in a straightforward way long gamma ray bursts (Detmers et al., 2008).
\medskip
\item	Rapidly rotating stars are formed via the mass transfer process in binaries or via the binary merging process (due to common envelope evolution or due to dynamics in dense star systems). This means that a cluster where the initial population consists mainly of slow rotators but with a significant population of binaries, will become populated with rapid rotators due to binary evolution or due to the interplay of binaries and cluster dynamics. At least part of the cluster Be-type star population is expected to be formed this way (Pols and Marinus, 1994; Van Bever and Vanbeveren, 1998).
\medskip
\item On of the hot items in stellar evolution research is the formation and evolution of stellar mergers. Mergers result as a consequence of canonical binary evolution (due to a non-conservative Roche lobe overflow process and/or due to common envelope evolution) or as a consequence of close encounters in star clusters where (in most cases) at least one of the players is a binary. It may therefore be expected that mergers are rapid rotators. Below I list 6 different kinds of mergers which have been studied in literature, but one may think of more combinations: 1. the merger of two main sequence (MS) stars. SPH simulations reveal that during the merger process large scale mixing happens (this means that in a cluster they will show up as blue stragglers) and mass is lost. As an example, Suzuki et al.  (2007) calculated the merging of two massive MS stars (a 88+88 M$_\odot$ and a 88+28 M$_\odot$ merger). After the merging process the new star is largely homogenized (which means that this star shows the products of CNO burning in the atmosphere) whereas during the merging process $\sim$10 M$_\odot$ is lost. It is tempting to link these results with the $\eta$ Car event in the 19$^{th}$ century. 2. The merger of a Wolf-Rayet star and a MS star. No detailed models have been calculated but it can be expected that the resulting star may be quite spectacular. 3.  The merger of a NS (BH?) with a MS star (a Thorne-Zytkow object, Thorne and Zytkow, 1977). Calculations of Canon et al. (1992) indicate that such objects may show up as red supergiants. Their further evolution is uncertain. 4. The merger of a WD and a MS star. Population synthesis of intermediate mass binaries reveal that many WD + MS binaries merge during the common envelope phase when the MS star fills its Roche lobe (De Donder and Vanbeveren, 2004). The further evolution of these mergers is also uncertain but I guess that the merger will be a rapid rotator and may show up as a Be star. 5. Double neutron star mergers, rapid rotators and a favorite model for short gamma ray bursts and possible sites of r-process element production and ejection (Dessart et al., 2009, see also section 5). 6. The merger of two WDs, a valuable model to explain SN Ia if one accounts properly for the effects of rotation during the merging process (Piersanti et al., 2003 and see section 4).
\medskip
\item	Accounting for the foregoing six points, the following statement is worth considering: Ôthe effect of rotation is important for the evolution of some massive stars but perhaps mainly in the framework of binaries or in the framework of binaries in combination with stellar dynamics in dense clustersÕ.
\medskip
\item	Binaries are an essential ingredient in population number/spectral synthesis. Examples: the evolution of massive star spectral features in starburst galaxies (Van Bever and Vanbeveren, 1998, 2003; Belkus et al., 2003; Brinchmann et al., 2008); the UV-upturn in elliptical galaxies (Han et al., 2007); the X-ray binary population in galaxies (Van Bever and Vanbeveren, 2000); the population of double pulsars (De Donder and Vanbeveren, 1998, 2003; Belczynski et al., 2002); the population of carbon enhanced metal poor stars (Pols et al., 2009); the population of short gamma ray bursts (= merging of double neutron star binaries) (De Donder and Vanbeveren, 1998; O'Shaughnessy et al., 2009); the population of long gamma ray bursts (Detmers et al., 2008) and last not least, the SN Ia (see section 4) which are responsible for some 70\% of all the iron in the universe.

\medskip
\item	The discovery of the double pulsar J0737-3039 and recent population synthesis models of massive binaries have reopened the discussion on the origin of r-process elements (De Donder and Vanbeveren, 2003) (see section 5).
\medskip
\item	Massive population III binaries where the primary is a very massive star which ends its life in a pair-instability supernova, may be important sites of primary nitrogen (Vanbeveren and De Donder, 2006)
\medskip
\item Most of the theoretical models that aim at explaining the chemical evolution of galaxies intrinsically assume that all stars are single stars (e.g., only single star yields are used). However, most of the massive stars are born as binary components and De Donder and Vanbeveren (2004) showed that the integrated chemical yields of a population of massive binaries differs by a factor 2-3 from the integrated chemical yields of a population of massive single stars. I think that it is time to include binaries in all galactic chemical codes. 
\end{itemize} 
 
\section{The Ultra Luminous X-ray source in the young dense cluster MGG11}
\medskip
Ultra Luminous X-ray sources (ULX) are point sources with X-ray luminosities up to 10$^{42}$ erg s$^{-1}$. MGG 11 is a young dense star cluster with Solar type metallicity $\sim$200 pc from the centre of the starburst galaxy M82, the parameters of which have been studied by McCrady et al. (2003). A ULX is associated with the cluster. When the X-rays are due to Eddington limited mass accretion onto a black hole (BH) it is straightforward to show that the mass of the BH has to be at least 1000 M$_\odot$. However how to form a star with Solar metallicity and with a mass larger than 1000 M$_\odot$? Mass segregation in a dense young cluster associated with core collapse and the formation of a runaway stellar collision process was promoted by Portegies Zwart et al. (2004). Note that the latter paper mainly addressed the dynamical evolution of a dense cluster but the evolution of the very massive stellar collision product was poorly described. 

The evolution of very massive stars has been studied in detail by Belkus et al. (2007) and it was concluded that stellar wind mass losses during core hydrogen burning and core helium burning are very important. A convenient evolutionary recipe for such very massive stars was presented, which can easily be implemented in an N-body dynamical code. Our N-body code which includes this recipe has been described in Belkus (2008) and in Vanbeveren et al. (2009) and applied in order to simulate the evolution of MGG 11. Our main conclusion was  the following:

\medskip
{\it \noindent Stellar wind mass loss of massive and very massive stars does not prevent the occurrence of a runaway collision event and the formation of a very massive star in a cluster like MGG 11, but after this event stellar wind mass loss during the remaining core hydrogen burning phase is large enough in order to reduce the mass again and the formation of a BH with a mass larger than $\sim$75 M$_\odot$ is rather unlikely.}

\medskip
The foregoing calculations and conclusion depend critically on the adopted stellar wind mass loss formalism for very massive stars. We used a formalism proposed by Kudritzki (2002) and this requires some discussion. Kudritzki published rates for stars with a luminosity up to log L/L$_\odot$ = 7, but we extrapolated his formalism for stars with a mass larger than 1000 M$_\odot$ corresponding to Log L/L$_\odot$ $\ge$ 7.5. Furthermore, the Kudritzki results are calculated with the Munich stellar wind code as it was in the year 2000 and in this code the line force was not yet calculated consistently with the line blocking/blanketing NLTE computations (Pauldrach, 2009 priv. communication). Using the new more consistent version of the Munich code (for a description see Pauldrach et al., 2003) stellar wind mass loss rates were determined for very massive stars with a mass up to 3000 M$_\odot$ using the properties of these stars predicted by evolution (Pauldrach and Vanbeveren, 2009). A main result is that the new mass loss rates are significantly smaller that the Kudritzki values. To illustrate, for a star with Teff = 50kK, M/M$_\odot$ = 250 and Log L/L$_\odot$ = 6.9, the old mass loss rate = 2.5 10$^{-4}$ M$_\odot$/yr, whereas the new value is 5.2 10$^{-5}$ M$_\odot$/yr which is about a factor five smaller. 

We implemented the new rates in our N-body code (note that our code combines consistently the effects of dynamics and life stellar evolution of very massive stars using a very efficient evolutionary algorithm) and we recalculated the dynamical evolution of MGG 11 using the same initial cluster conditions as in Vanbeveren et al. (2009).  During core helium burning (the WR phase) of the very massive stars we used the same WR wind formalism as in the latter study. The new mass evolution of the runaway merger is illustrated in figure 1. We notice a similar behavior as in the old simulations, e.g. a rapid mass growth that is not prevented by stellar wind mass loss and significant mass loss after the main collision event, but as expected, the final mass is significantly larger. The formation of a IMBH in the cluster MGG 11 with a mass between 200 M$_\odot$ and 300 M$_\odot$ cannot be excluded. On the question what is the maximum mass of the BH formed dynamically in a cluster like MGG 11, the final answer my friend is clearly blowing in the stellar wind mass loss of very massive stars.

\begin{figure}[b]
\begin{center}
\includegraphics[width=5in]{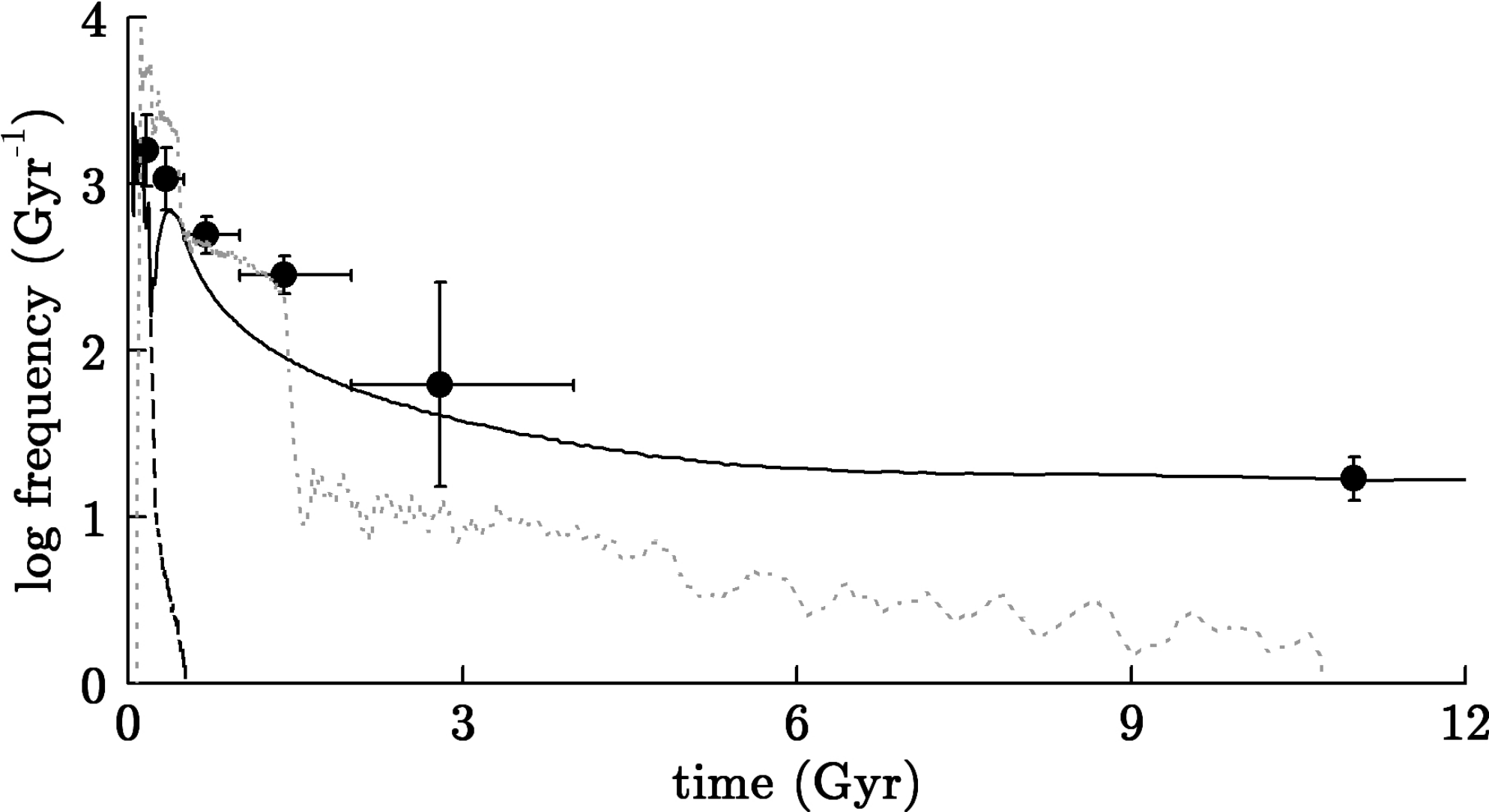} 
 \caption{Delayed time distributions (DTDs) for $\beta$ = 1 in case of the DD (solid black) and the SD (dotted grey) scenario as well as DTD for $\beta$ = 0 in case of the DD (dashed black) scenario. Observational data points of Totani et al. (2008).}
\label{fig1}
\end{center}
\end{figure}

\section{The delay time distribution of SN Ia}
\medskip
It is generally believed that Type Ia supernovae (SN Ia) are thermonuclear explosions of white dwarfs (WDs) that exceed their Chandrasekhar limit. There are two main scenarios explaining how this can happen, and both involve the evolution of interacting intermediate mass binaries: the single degenerate scenario (SD) and the double degenerate scenario (DD). In the SD scenario, the WD has a main sequence or a red giant companion that is filling its Roche lobe. Hydrogen rich matter is transferred towards the WD and this pushes the WD mass over the Chandrasekhar limit. The DD model involves the formation of a double CO WD binary. Due to gravitational wave radiation both WDs spiral-in and merge. When the mass of the merger is larger than 1.4 M$_\odot$, a SN Ia may happen (Webbink, 1984). It has been argued that the merging of two WDs will lead to the formation of a neutron star and a SN Ia will not happen (e.g., Saio and Nomoto, 1998 and references therein). However, these contra-arguments do not account for the consequences of rotation and angular momentum transport during the merging process. This effect was investigated by Piersanti et al. (2003) (see also more recent  papers published by this group) and they showed that the rotating DD model produces in a natural way a hydrogen less SN Ia. 

In order to solve the question which of the two scenarios is the dominant contributor to the SN Ia rate, population synthesis of starburst regions may be very useful and we focus on the delay time distribution (DTD) of SN Ia. The DTD is defined as the number of SN Ia events as a function of time in a starburst. By observing elliptical galaxies, which are for this purpose equivalent to starburst galaxies, at similar metallicity but at different redshifts, it is possible to construct an observational DTD. This can then be compared to DTDs for starburst galaxies predicted by population synthesis. Studies like that have been performed by Yungelson and Livio (2000, only DD models), Han and Podsiadlowski (2004, only SD models) and Ruiter et al. (2008, SD and DD models).  We have recently performed a similar study (Mennekens et al., 2009) and I summarize some of the results here.  

The theoretical DTDs are calculated by using an updated version of the population code of De Donder and Vanbeveren (2003) and we compare with the observational DTD of Totani et al. (2008). We obviously account for the common envelope process in binary evolution but we also focus on the evolution of Case A and Case Br binaries that evolve via a canonical Roche lobe overflow (RLOF) process, mass transfer and mass accretion (as observed in Algol systems). Binary population synthesis results depend on many parameters and after a detailed parameter study we conclude that

\medskip
{\it \noindent Double WD progenitors experience two RLOF phases. The RLOF may result into the formation of a common envelope, however, our population predictions reveal that the first RLOF of most of the DD SN Ia progenitors is a canonical RLOF with mass transfer and mass accretion (the second RLOF when the original primary is already a WD is obviously a common envelope process). }
\medskip

An important consequence of the conclusion above is that any analytical formalism in order to describe the DTD of DDs that is based on the assumption that the progenitors went through two common envelope phases, is wrong.
 
Figure 2 shows the DTD prediction of the DD model in case that the first RLOF of the progenitor binaries is assumed to be conservative (all mass lost by the loser is accreted by the gainer) and non-conservative (all mass lost by the loser leaves the system through the second Lagrangian point). Comparison with the observed DTD allows us to conclude that only a population model where the first RLOF is (quasi)-conservative gives reasonable correspondence with observations. Notice that this conclusion corresponds with the conclusion when population synthesis predictions of Algol binaries is compared with the observed properties of Algol binaries in the Solar Neighborhood. 
 
In figure 2 we also show the DTD of the SD model using the SD progenitors as given by Hachisu et al. (2008). In the latter paper these progenitors are identified as contours for different WD mass, as a function of binary period and companion mass. A SD SN Ia is assumed to result if the evolutionary track of the progenitor system traverses this contour. As can be noticed especially the late time behavior of the DTD gives poor correspondence with observations. 
Overall conclusion  

\medskip
{\it \noindent the predicted delayed time distribution with single degenerate SN Ia progenitors only does not reproduce the late time behavior of the observed DTD. The predicted DTD with double degenerate SN Ia progenitors (eventually in combination with SD progenitors) and a (quasi-)conservative Roche lobe overflow in Case A and Case Br intermediate mass binaries gives the best correspondence.} 

\section{The delay time distribution of NS+NS and NS+BH mergers}
\medskip
It is generally accepted that the rapid neutron-capture process (r-process) is responsible for the existence of the heaviest elements in the universe. There are two sites where the physical conditions are such that the r-process can happen: the SN explosion of a massive star and the binary neutron star merger (NS+NS and possibly also NS+BH) (Qian and Woosley, 1996; Rosswog et al., 2001; Dessart et al., 2009).

\begin{figure}[b]
\begin{center}
\includegraphics[width=4in]{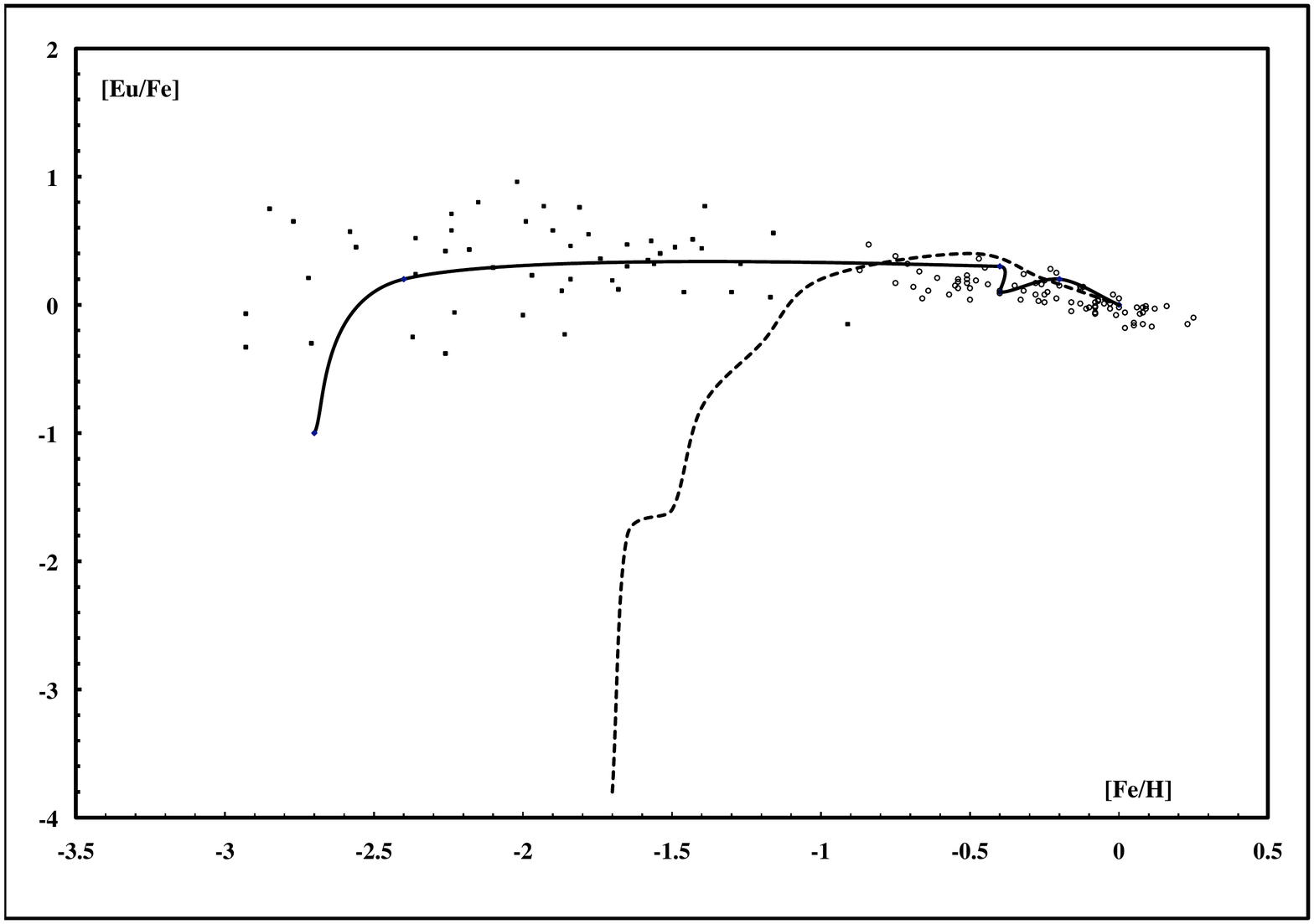} 
 \caption{The temporal evolution of the predicted double neutron star merger rate in the Solar Neighborhood. The dotted curve corresponds to the one predicted by Matthews (1992) whereas the full line is the one of De Donder and Vanbeveren (2003, 2004). The observations are from various sources discussed in the latter two papers.}
\label{fig1}
\end{center}
\end{figure}

\begin{figure}[b]
\begin{center}
\includegraphics[width=4in]{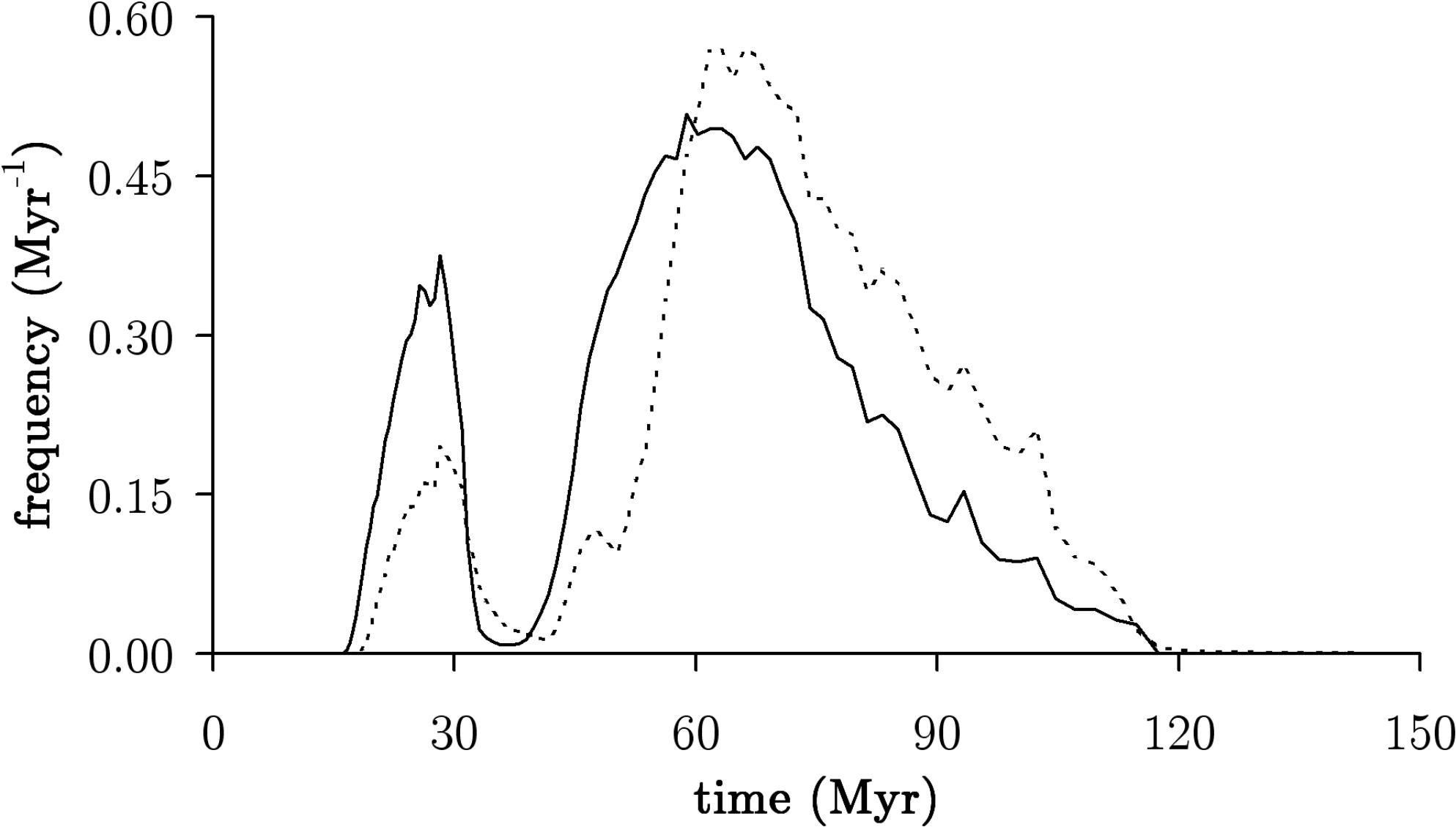} 
\caption{Delayed time distributions of double neutron star mergers (DTDmergers) using standard population synthesis parameters (see text). The full curve corresponds to Z = 0.02, the dotted one to Z = 0.002.}
\label{fig1}
\end{center}
\end{figure}

One of the major arguments against the binary neutron star merger has been published by Matthews et al. (1992). He considered the observed periods and eccentricities of the binary neutron stars known at that time and calculated the expected merger rate time scale using the theory of gravitational wave radiation. He implemented this in a galactic evolutionary model and when a comparison was made between the merger rate and the observed galactic evolution of the r-process element europium, he concluded that compared to Eu double compact star mergers appear too late (see the dashed line in figure 3) and thus, neutron star binary mergers cannot be major r-process production sites. However, the discovery of the short period (2.4 hours) double pulsar J0737-3039 (Lyne et al., 2004) reopened the discussion. Moreover, since 1992 our physics view of the evolution of binaries improved a lot. Ivanova et al. (2003) and Dewi and Pols (2003) studied the Case BB evolution of binaries with a hydrogen deficient helium burning component and a normal main sequence companion. They proposed a scenario which leads to the formation of ultra-compact double neutron star binaries which can merge on a 1 million year time scale. De Donder and Vanbeveren (2003, 2004) included this scenario in a binary population synthesis model and calculated the delayed time distribution of these mergers (DTDmerger), defined as the evolution of double neutron star mergers in a starburst. The various evolutionary phases of massive binaries prior to double neutron star binary formation depends on a number of uncertain parameters and also the {DTDmerger} depends on these parameters. A typical distribution is shown in figure 4 for Z = 0.02 and 0.002 (Salpeter type IMF, a binary mass ratio distribution that is flat, the binary period distribution is flat in the log, common envelope evolution efficiency $\alpha$ = 1, mass transfer efficiency during the canonical Roche lobe overflow phase in Case Br binaries $\beta$ = 1, our preferred kick velocity distribution during the SN explosion with an average v$_{kick}$ = 450 km/s). As can be noticed, double neutron star mergers appear very early. De Donder and Vanbeveren included this model in a galactic chemical code and figure 3 shows that the temporal evolution of the average merger rate now follows the observed evolution of Eu in the Solar Neighborhood.

\section{The role of massive binaries in the self-enrichment process of globular clusters}
\medskip
It is generally accepted that a large fraction of the low mass stars in globular clusters (GCs) is formed from material that was enriched with hydrogen burning products produced in more massive stars: the self-enrichment process in GC. 

The first self-enrichment scenario that was proposed in literature was the one in which the enrichment was due to low metallicity (Z) intermediate mass stars (Cottrell and Da Costa, 1981). Detailed intermediate mass evolutionary calculations manage to explain the observed abundance patterns in GCs but fine tuning is required of the evolutionary processes, especially those that operate during the AGB phase (e.g., Decressin et al., 2009; Ventura and DÕAntona, 2009). Note that when the GC initially had an intermediate mass binary population similar to the one in other aggregates (like in the association Sco OB2 discussed by Kouwenhoven, 2006), and when the DD scenario (Webbink, 1984) is responsible for the SN Ia events, population synthesis reveals that many SN Ia should have happened in the past. To illustrate, using the Brussels population code, a starburst simulation with 10$^6$ M$_\odot$ of stars with initial mass $\ge$ 0.8 M$_\odot$ and where 30 \% of the intermediate mass stars are primaries of binaries with an orbital period $\le$ 10 years, predicts 10000-15000 SN Ia events. In order not to conflict with the observed Z of GCs, it is clear that most of the matter ejected during the SN Ia must have left the cluster. Interestingly, if the SD scenario applies, then we do not expect many SN Ia at all. The reason is that the number of SN Ia predicted by the SD scenario strongly depends on Z, low Z implies very few SN Ia (Hachisu et al., 2008).

A second self-enrichment scenario is based on the assumption that prior to the formation of the low mass stars, a population of low-Z massive stars was present. This model may work if a process is available in order to remove hydrogen burning products at low velocity (smaller than the escape velocity of the cluster) from the low-Z massive stars. Decressin et al. (2007) propose the {\it Winds of Fast Rotating Massive Stars} scenario. For this scenario to work, fast means really very fast (equatorial velocities of the order 800-1000 km/s) and it remains to be demonstrated if such high average values are real in small metallicity regions. However, when a massive star is a binary member it may lose its CNO-processed layers in a natural way by the RLOF/common envelope/spiral-in process and fast rotation is not needed (remind that most of the massive stars in our Galaxy are observed as binary members, section 2). So, I like to propose the {\it RLOF/common envelope/spiral-in Mass Loss in Massive Binaries} scenario. 

The delayed time distribution of merging double neutron star binaries has been discussed in section 5 and it was concluded that the galactic temporal evolution of these mergers follows the observed temporal evolution of Eu in the Solar Neighborhood, which may be an indication the they are production sites of r-process elements that cannot be neglected.  I suggest that these binaries may also be important sites of r-process self-enrichment of GCs.

\end{document}